\documentclass[prb,twocolumn,amsmath,showpacs]{revtex4-1}
\usepackage{bm}
\usepackage{graphicx}

\newcommand{\ket}[1]{|#1\rangle}
\newcommand{\bracket}[1]{\langle #1 \rangle}
\newcommand{\up}{\uparrow}
\newcommand{\dn}{\downarrow}
\newcommand{\eps}{\varepsilon}

\begin{document}

\title{Optical generation and detection of pure valley current in monolayer transition metal dichalcogenides}

\author{Wen-Yu Shan}
\affiliation{Department of Physics, Carnegie Mellon University, Pittsburg, Pennsylvania 15213, USA}
\author{Jianhui Zhou}
\affiliation{Department of Physics, Carnegie Mellon University, Pittsburg, Pennsylvania 15213, USA}
\author{Di Xiao}
\affiliation{Department of Physics, Carnegie Mellon University, Pittsburg, Pennsylvania 15213, USA}
\date{\today}

\begin{abstract}
We propose a practical scheme to generate a pure valley current in monolayer transition metal dichalcogenides by one-photon absorption of linearly polarized light.  We show that the pure valley current can be detected by either photoluminescence measurements or the  ultrafast pump-probe technique.  Our method, together with the previously demonstrated generation of valley polarization, opens up the exciting possibility of ultrafast optical-only manipulation of the valley index. The tilted field effect on the valley current in experiment is also discussed.
\end{abstract}

\maketitle

\section{Introduction}

Recent years have seen a surge of interest in the manipulation of the valley index of Bloch electrons,~\cite{gunawan2006,rycerz2007,xiao2007,yao2008,gunlycke2011,zhu2012,jiang2013,isberg2013}
largely driven by its potential applications in electronics and optoelectronics.~\cite{natphys}
The valley index enumerates degenerate energy extrema in momentum space. Such degeneracies are often present in 2D materials with a honeycomb-like structure, such as graphene, boron nitride, and transition metal dichalcogenides (TMD).
In these materials, weak intervalley scattering renders the valley index an effective degree of freedom that can be utilized in novel devices.
This realization of valley-based electronics is called valleytronics, which depends crucially on the \emph{dynamical} control of two quantities: valley polarization and valley current.
The optical generation of valley polarization by circularly polarized light~\cite{xiao2012,cao2012} shows some promise as a realization of valleytronics and has been experimentally demonstrated in monolayer MoS$_2$.~\cite{zeng2012,mak2012,cao2012} On the other hand, so far only valley-polarized electric current has been reported.~\cite{mak2014,yuan2014,zhang2014a} In analogy to spintronics, it would be desirable to generate a pure valley current, in which there is no net motion of charge; carriers in opposite valley move in opposite direction. Such a pure valley current would rule out any charge-related effect~\cite{jedema2001,jedema2003,yang2008} and generate minimal Joule heating, similar to a pure spin current~\cite{murakami2003}.

In this work, we propose a new approach to the generation and detection of a pure valley current by optical means.
Based on both symmetry analysis and an effective $\bm k\cdot \bm p$ Hamiltonian, we show that a pure valley current can be generated by linearly polarized light in monolayer TMDs.
The generating mechanism parallels that for spin current.~\cite{bhat2005}
However, the role of spin-orbit coupling is replaced by the trigonal warping in the band structure, which is entirely a lattice effect.
Due to the unique spin-valley coupling in this system,~\cite{xiao2012} the generated valley current is accompanied by a spin current.
We also present a theory for valley diffusion that takes into account the spin-valley coupling, and show that the pure valley current can be detected by either photoluminescence measurements or the  ultrafast pump-probe technique.
Our method, together with the previously demonstrated generation of valley polarization,~\cite{zeng2012,mak2012,cao2012} opens up the exciting possibility of ultrafast optical-only manipulation of the valley index.

The paper is organized as follows. In Sec. \ref{sec:generation}, we present the optical generation of valley current, where the symmetry analysis is given in Sec. \ref{sec:symmetry} and the numerical result is shown in Sec. \ref{sec:numerical}. Detection of the generated valley current is considered in Sec. \ref{sec:detection}, where the ultrafast pump-probe and photoluminescence measurements are proposed in Sec. \ref{sec:ultrafast} and Sec. \ref{sec:pl}, respectively. Finally, discussion and conclusion are made in Sec. \ref{sec:discussion}.

\section{\label{sec:generation}Optical generation of valley current}

\subsection{\label{sec:symmetry}Symmetry analysis}

\begin{figure}[t]
\centering \includegraphics[width=0.45\textwidth]{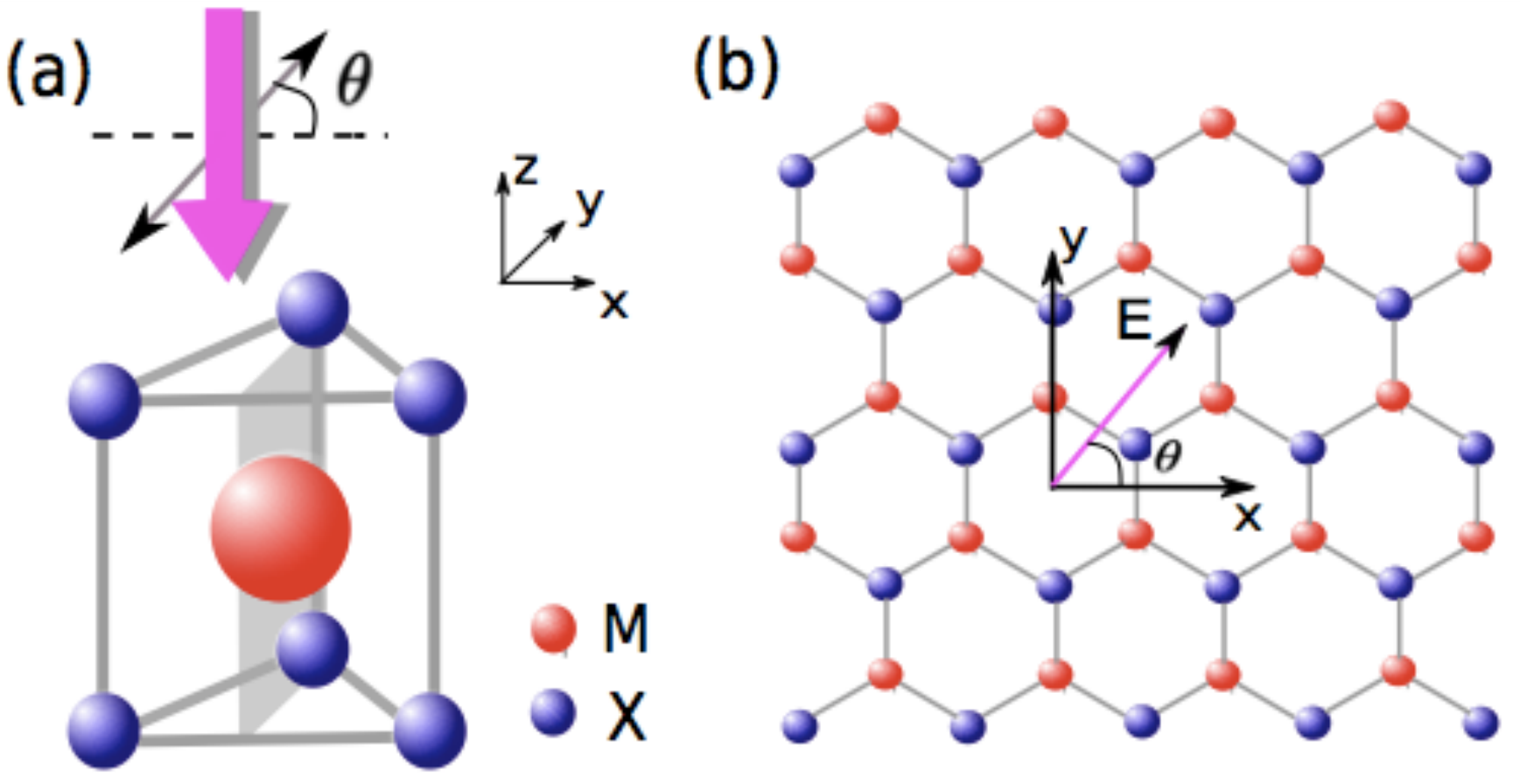}
\caption{(color online). The crystal structure of monolayer TMD, where $M$ (red) is the transition metal atom and $X$ (blue) is the chalcogen atom.  (a) Side and (b) top view of the lattice structure.  The thick arrow depicts a normally incident linearly polarized light with the polarization angle $\theta$. The $yz$ plane is defined as one of the mirror planes (shaded region). }
\label{fig:lattice}
\end{figure}

Figure~\ref{fig:lattice} shows the schematic setup.
A linearly polarized light $\bm E(\omega) = E_0(\cos\theta\hat{\bm x} + \sin\theta\hat{\bm y})$ at normal incidence is considered, where $E_0$ and $\theta$ refer to the amplitude and polarization angle of the electric field, respectively.
We choose $\hat{\bm x}$ to be along the zigzag direction and $\hat{\bm y}$ the armchair direction.
In monolayer TMD, each transition metal cation is trigonal-prismatically coordinated by six nearest neighbor chalcogen anions, explicitly breaking the inversion symmetry.
The relevant symmetry operations of the system include three-fold rotation $C_3$ around the $\hat{\bm z}$ axis, mirror reflections $M_x(x \rightarrow -x)$ and $M_z(z \rightarrow -z)$, and time-reversal.

The generation of a dc current by imposing an optical field---namely, the photogalvanic effect (PGE)---is a second-order nonlinear phenomenon characteristic of non-centrosymmetric materials.
Under a monochromatic light $\bm E(t)=\bm E(\omega) e^{-i\omega t} + \text{c.c.}$, the photocurrent has the standard form
\begin{equation}
\begin{split}\label{definition}
j_{\alpha}&= \sum_{\bm k}\chi_{\alpha\beta\gamma}(\bm k,\omega,-\omega)E_{\beta}(\omega)E_{\gamma}^*(\omega) \;,
\end{split}
\end{equation}
where the $\bm k$-resolved second-order susceptibility tensor $\chi_{\alpha\beta\gamma}(\bm k,\omega,-\omega)$ is given from the perturbation theory by~\cite{kraut1979,boyd2008}
\begin{equation}
\begin{split}\label{definition2}
\chi_{\alpha\beta\gamma}(\bm k,\omega,-\omega)&=\frac{e^3}{\hbar^2\omega^2S}\int_{-\infty}^0dt_1\int_{-\infty}^{t_1}dt_2e^{-i\omega(t_1-t_2)}\\
&\times e^{t_2/\tau}\chi_{\alpha\beta\gamma}(\bm k,t_1,t_2), \\
\chi_{\alpha\beta\gamma}(\bm k,t_1,t_2)&=\mathrm{Tr}(\hat{\rho}_0(\bm k)[[\hat{v}_{\alpha},\hat{v}_{\beta}(t_1)],\hat{v}_{\gamma}(t_2)]).
\end{split}
\end{equation}
$\hat{\rho}_0(\bm k)$ is the initial equilibrium density matrix operator and $\hat{v}_{\alpha,\beta,\gamma}(t)$ are the velocity operator in the Heisenberg picture at time $t$. $\mathrm{Tr}$ denotes the trace and $S$ is the area. Since we are dealing with a strictly 2D system, the indices $\alpha,\beta,\gamma$ can be either $x$ or $y$. According to Ref. \onlinecite{hosur2011}, $\chi_{\alpha\beta\gamma}(\bm k,\omega,-\omega)$ is further decomposed into three terms: two constant terms and one linear-in-time term, the latter of which is cut off by relaxation time $\tau$ based on the relaxation time approximation. In this work, we restrict ourselves to the high quality samples with $\tau\gg\hbar/\Delta E$, in which the response is dominated by the linear-in-time term, and the other two terms can be neglected. \footnote{In low-quality sample or system with small optical transition gap, the linear photogalvanic effect is non-negligible and valley current becomes partially polarized. This may explain the observed photocurrent in Ref. \onlinecite{yuan2014}. } $\Delta E$ is the optical transtion gap. As a result, the susceptibility reduces to
\begin{equation}
\begin{split}\label{definition3}
\chi_{\alpha\beta\gamma}(\bm k,\omega,-\omega)&= -\frac{\pi e^3\tau}{\hbar\omega^2S}\sum_{n,m}(v_{\beta,\bm k})_{nm}(v_{\gamma, \bm k})_{mn}\\
&\times[(v_{\alpha, \bm k})_{mm}-(v_{\alpha,\bm k})_{nn}]F_{nm,\bm k}\\
&\times\delta(\hbar\omega+\eps_{m\bm k}-\eps_{n\bm k}) \;.
\end{split}
\end{equation}
Here $(v_{\alpha,\bm k})_{mn} = \bracket{m,\bm k|\hat v_\alpha|n,\bm k}$ is the velocity matrix element in the Bloch basis $\ket{n,\bm k}$, $\eps_{n\bm k}$ is the band dispersion, $F_{nm,\bm k}=f_{n\bm k} - f_{m\bm k}$ with $f_{n\bm k}=[1+\exp[\beta(\epsilon_{n,\bm k}-\mu)]]^{-1}$ being the Fermi-Dirac distribution. $\mu$ is the chemical potential and $\beta=1/k_BT$. Note that the expression for $j_{\alpha}$, when transformed into the real space representation, is also recognized as the ``shift current''.~\cite{sturman1992,sipe2000}

For a system with time-reversal symmetry (TRS), PGE vanishes under linearly polarized light.
The reason is that linear polarization picks out the real part of $\chi$ in Eq.~\eqref{definition}, which satisfies
\begin{equation}
\label{schi}
\chi^{\up}_{\alpha\beta\gamma}(\bm k)=-\chi^{\dn}_{\alpha\beta\gamma}(-\bm k)\;, \quad
\chi^{\dn}_{\alpha\beta\gamma}(\bm k)=-\chi^{\up}_{\alpha\beta\gamma}(-\bm k) \end{equation}
due to the TRS.
For simplicity, we have omitted the arguments $\omega$ and $-\omega$ in $\chi$.
Summing over each pair of $\chi$'s in Eq.~\eqref{schi} then yields zero charge current.
Equation~\eqref{schi} suggests that it is possible to generate a pure spin current.
However, without breaking the spin degeneracy, one has $\chi^{\up}_{\alpha\beta\gamma}(\bm k)=\chi^{\dn}_{\alpha\beta\gamma}(\bm k)$; consequently, the total spin current still vanishes.
In Ref.~\onlinecite{bhat2005}, Bhat~\textit{et al.}\ showed that introducing the spin-orbit coupling can break the spin degeneracy, giving rise to a pure spin current.

We now show that a similar effect can generate a pure valley current, i.e., $j^K+j^{K^{'}}=0$, and $j^K\neq0$.
In monolayer TMD, the two valleys, located at the $K$ and $K'$ points of the hexagonal Brillouin zone, are related by the TRS.
As such, the valley-resolved susceptibility tensor satisfies
\begin{equation}
\chi^{K}_{\alpha\beta\gamma}(\bm q)=-\chi^{K'}_{\alpha\beta\gamma}(-\bm q) \;, \quad
\chi^{K'}_{\alpha\beta\gamma}(\bm q)=-\chi^{K}_{\alpha\beta\gamma}(-\bm q) \;,
\end{equation}
where $\bm{q} \equiv \bm k-\bm K(\bm K')$ defines a small momentum away from the valley center $K$ ($K')$.  Again, the charge current vanishes.  However, the $K$ and $K'$ points have $C_3$ rotational symmetry. This allows $\chi^{K}_{\alpha\beta\gamma}(\bm{q})\neq\chi^{K'}_{\alpha\beta\gamma}(\bm{q})$, leading to a pure valley current $\propto\sum_{\bm{q}}\sum_{\eta}\xi_{\eta}\chi^\eta_{\alpha\beta\gamma}(\bm{q})$,where $\xi_{\eta}=\pm1$ for $\eta=K,K'$.
For a system symmetric under $M_z$, $C_3$ is the only possible rotation symmetry that can break the valley ``degeneracy'' and induce the valley current;
all other rotational symmetries yield zero valley current, even though they given rise to an anisotropic band structure.

\begin{figure}[t]
\centering \includegraphics[width=0.30\textwidth]{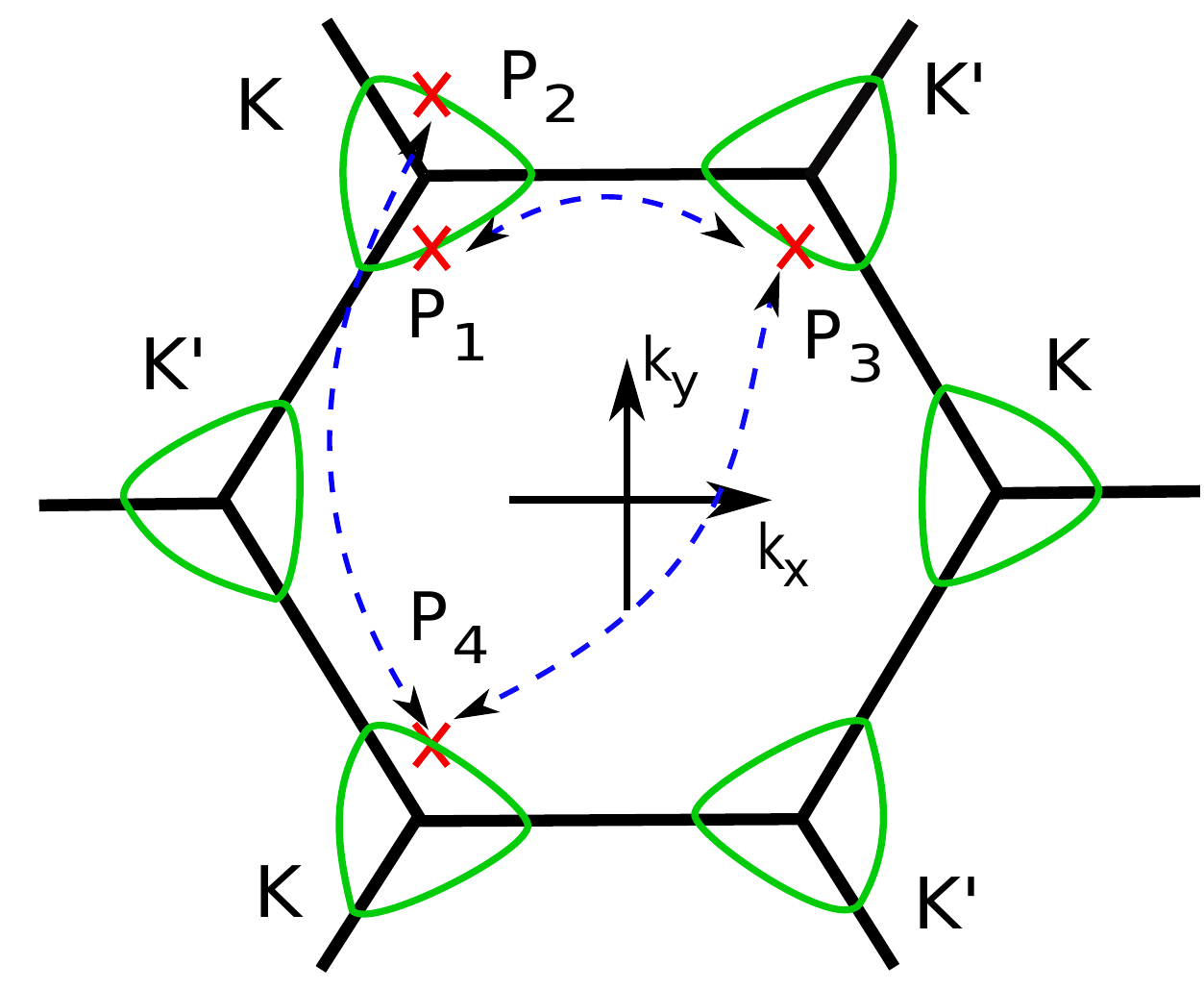}
\caption{(color online). The first Brillouin zone (BZ) of monolayer MX$_2$. $P_{1,2,3,4}$ are symmetry-related points. }
\label{fig:BZ}
\end{figure}

To substantiate the preceeding argument, we carry out a detailed group theory analysis.  The symmetry group of the $K$ point is $C_{3h}$;  $\chi$ transforms as a direct product $E'\otimes E'\otimes E'$, which contains two copies of the identity representation.
This indicates that there are two independent components of $\chi$: $\chi_{yyy}^\eta=-\chi_{xxy}^\eta=-\chi_{xyx}^\eta=-\chi_{yxx}^\eta$ (denoted by $\chi_e^\eta$) and $\chi_{xxx}^\eta=-\chi_{xyy}^\eta=-\chi_{yxy}^\eta=-\chi_{yyx}^\eta$ (denoted by $\chi_o^\eta$).
However, due to the TRS, we find that the combined symmetry $TM_x$ requires the contribution from $\chi_e^\eta$ to vanish.
To see this, let us consider two $\bm k$-points $P_1$ and $P_2$ in the $K$ valley. These are related by $q_y \to -q_y$ (Fig.~\ref{fig:BZ}).
We also introduce two intermediate points $P_3$ and $P_4$. $P_1$ and $P_3$ are related by the mirror symmetry $M_x$ and satisfy
\begin{align}
\chi^K_e(q_x,q_y) &= \chi^{K'}_e(-q_x,q_y) \;.
\end{align}
Meanwhile, $P_3$ and $P_4$ are related by the TRS, so that
\begin{align}
\chi^K_e(q_x,q_y)&=-\chi^{K'}_e(-q_x,-q_y) \;.
\end{align}
Finally, $P_2$ and $P_4$ are equivalent up to a reciprocal lattice vector.
Therefore, upon summing over $\bm k$-states in one valley, the contribution from $\chi_e^\eta$ vanishes.
The only non-vanishing contribution to the valley current is from $\chi_o^\eta$, with an angular dependence
\begin{equation}
j^\eta \propto \chi^\eta_{xxx} \cos(2\theta+\varphi) \;,
\end{equation}
where $\varphi$ is the detection angle.
Experimentally, by fixing $\varphi$, a $\pi$-period oscillation of the signal is expected.

\subsection{\label{sec:numerical}Numerical results}

\begin{figure}[t]
\centering \includegraphics[width=0.50\textwidth]{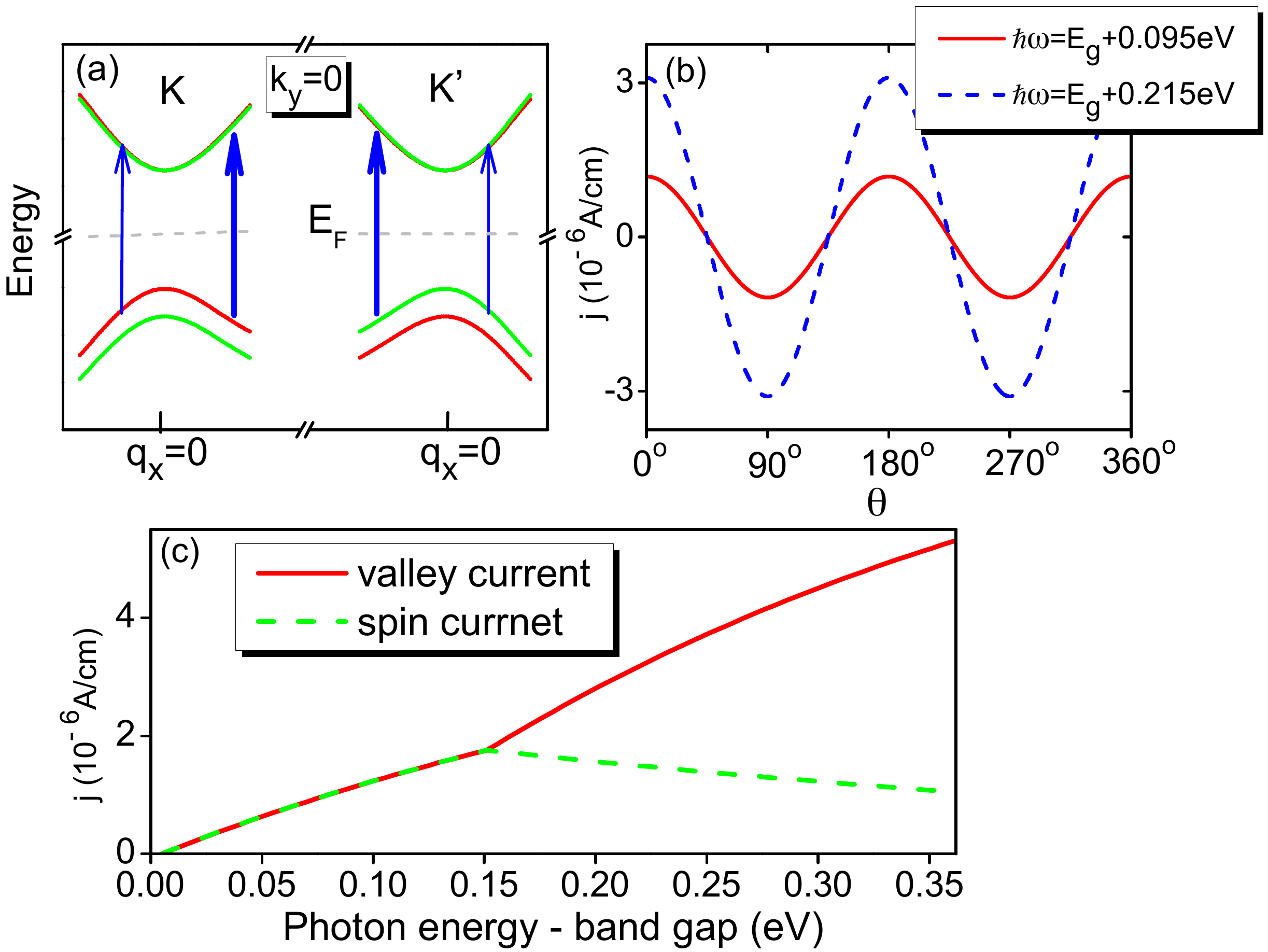}
\caption{(color online). Linear valley and spin photogalvanic effect (PGE) of monolayer MoS$_2$. (a) Schematics of band dispersion around valley $K(K')$ along the $k_x$ axis. Red (green) curves denote states with spin up (down). Thick (thin) solid blue arrows depict strong (weak) optical transition rates in each valley. Fermi energy $E_F=0$. (b) Angular dependence of valley current on the polarization angle $\theta$ along the $\hat{\bm x}$ axis (zigzag direction). Red solid and blue dashed curves label photon energy $\hbar\omega=1.68$eV and $1.80$eV. (c) Valley and spin current as functions of photon energy (minus by band gap $E_g$). $\tau=55$ fs, $E=3.01\times10^4$ V/m, $T=5$ K. Band parameters are adopted from Ref. \onlinecite{liu2013} and band gap $E_g=1.585$eV.}
\label{fig:LPGE}
\end{figure}

A unique property of monolayer TMD is the strong spin-valley coupling, which refers to the opposite spin splitting at the valence band edge in opposite valleys [see Fig.~\ref{fig:LPGE}(a)].~\cite{xiao2012}
We can see immediately that a pure spin current will accompany the generated valley current.
Note that this spin current would vanish if the energy bands at the $K$ and $K'$ points are isotropic, even in the presence of spin-orbit coupling.

Another important parameter in Eq.~\eqref{definition} is the relaxation time $\tau$.  Due to its multivalleyed band structure, there are several scattering channels in monolayer TMD. They give rise to four relaxation times $\tau^{e/h}_{intra/inter}$ which refer to intra- and intervalley scattering by electrons ($e$) and holes ($h$).  These lifetimes satisfy~\cite{lu2013,song2013,shan2013}
\begin{align}\label{timescale}
\tau^e_{intra}\sim\tau^h_{intra}\ll\tau^e_{inter}\ll\tau^h_{inter} \;.
\end{align}
Given that $1/\tau=1/\tau_{intra}+1/\tau_{inter}$, the optically generated valley current is predominantly determined by the intra-valley scattering time $\tau_{intra}$.
Electrons and holes exhibit almost the same intra-valley scattering time when neglecting the weak intrinsic electron-hole asymmetry in the system.
The last inequality in Eq.~\eqref{timescale} comes from the aforementioned spin-valley coupling.~\cite{lu2013,song2013,shan2013}  Although not essential in the generation of the valley current, it is important to the detection process as discussed below.
In the absence of spin-dependent scattering, the upper limit of $\tau^h_{inter}$ is set by the Bir-Aronov-Pikus mechanism, which could be as large as $\sim$ 1ns.~\cite{mak2012,ochoa2013}

To calculate the valley current in monolayer TMD, we employ a low-energy effective $\bm k\cdot \bm p$ Hamiltonian~\cite{liu2013} around valley $K (K')$, which includes both the $C_3$ symmetry-allowed trigonal warping and $k$-cubed corrections.  We also take into account the large spin splitting in the valence bands; the small spin splitting in the conduction bands is ignored.

With realistic parameters, \footnote{Let us consider a laser beam with power $I=100$ $\mu$W and the light spot radius $r=1$ $\mu$m. We assume an absorption coefficient $\xi=2\%$.~\cite{mak2013,liu2014} By making use of the formula $\xi P=\frac{1}{2}\epsilon_0\sqrt{\epsilon_r}cE^2$ with vacuum dielectric constant $\epsilon_0$ and in-plane relative permittivity $\epsilon_r=2.8$,~\cite{cheiwchanchamnangij2012} we obtain the electric field $E = 3.01\times10^4$ V/m. In addition, the mobility of monolayer MoS$_2$ is $\mu=200$ cm$^2$V$^{-1}$s$^{-1}$,~\cite{radisavljevic2011} which gives the momentum relaxation time $\tau=\mu m/|e|=55$ fs with $m$ being the effective mass.} our numerical results are shown in Fig.~\ref{fig:LPGE}.
In Fig.~\ref{fig:LPGE}(a), the band dispersion of the effective model is plotted for $k_y=0$, which clearly shows the large spin splitting in the valence bands.
Due to the  $C_3$ symmetry, optical transitions excite states with different $|q_x|$ in each valley.
This results in different optical transition rates (indicated by the thickness of the arrow) and different velocity in the $q_x$ direction, both of which contribute to generating the valley photocurrent.
Figure~\ref{fig:LPGE}(b) shows the angular dependence of the valley current on the polarization angle $\theta$ by fixing $\varphi=0$.
The valley current $j^v=j_{K}-j_{K'}$ has an order of $10^{-6}$ A/cm, comparable to the magnitude of photocurrent observed in GaAs quantum wells.~\cite{ganichev2003}
Figure~\ref{fig:LPGE}(c) displays both valley and spin current as functions of photon energy.
We note that as soon as the lower spin-split valence band becomes active, the spin current displays a downward trend.
This allows us to manipulate the generation of valley and spin current either collectively or separately.

\section{\label{sec:detection}Detection of valley current}

Next we discuss the detection of the pure valley current.
Our idea utilizes the fact that the valley carriers in monolayer TMD are described by a pair of massive Dirac fermions with opposite mass,~\cite{xiao2012} therefore each valley exhibits opposite time-reversal symmetry breaking effects such as circular dichroism~\cite{xiao2012} and Faraday rotation.~\cite{yang2013}
Note that there is a possible complication due to the large exciton binding energy observed in monolayer TMD,~\cite{zhang2014,chernikov2014,zhu2014,wang2014,he2014} which makes the generation of free carriers difficult.
To remove the exciton effect, we may heavily dope the sample~\cite{xu2014} or apply a large in-plane electric field.~\footnote{
The in-pane field could also induce a valley current proportional to the anomalous velocity.~\cite{xiao2007,xiao2012,mak2014}
To distinguish it from the valley photocurrent, one can use the property that the former (latter) is odd (even) under the reversal of the electric field.}
Under these circumstances, we propose two possible detection schemes.

\subsection{\label{sec:ultrafast}Ultrafast pump-probe measurement}

In one scheme, one can observe the second-harmonic generation (SHG) of the valley current using the ultrafast pump-probe technique~\cite{werake2010} as shown in Fig.~\ref{fig:detection}(a).
This is similar to the proposed detection method of spin current.~\cite{wang2010}
First, the pump light (with frequency $f_1 > E_g$) generates a pure valley current. Then the probe light (with frequency $f_2 < E_g/2$) creates a population imbalance between $\bm k$ and $-\bm k$ states, which leads to a net Faraday rotation.  This will induce a polarized field $\bm P(2f_2)\propto(\hat{\bm z}\times\bm E|\bm E|)$, and emit a second-harmonic signal (orthogonal to the probe light).
Since the energy of the SHG is still below the band gap $E_g$, the Faraday rotation is related to the virtual interband transition, which distinguishes it from other optical effects of the pump light.

\subsection{\label{sec:pl}Photoluminescence measurement}

\begin{figure}[t]
\centering \includegraphics[width=0.50\textwidth]{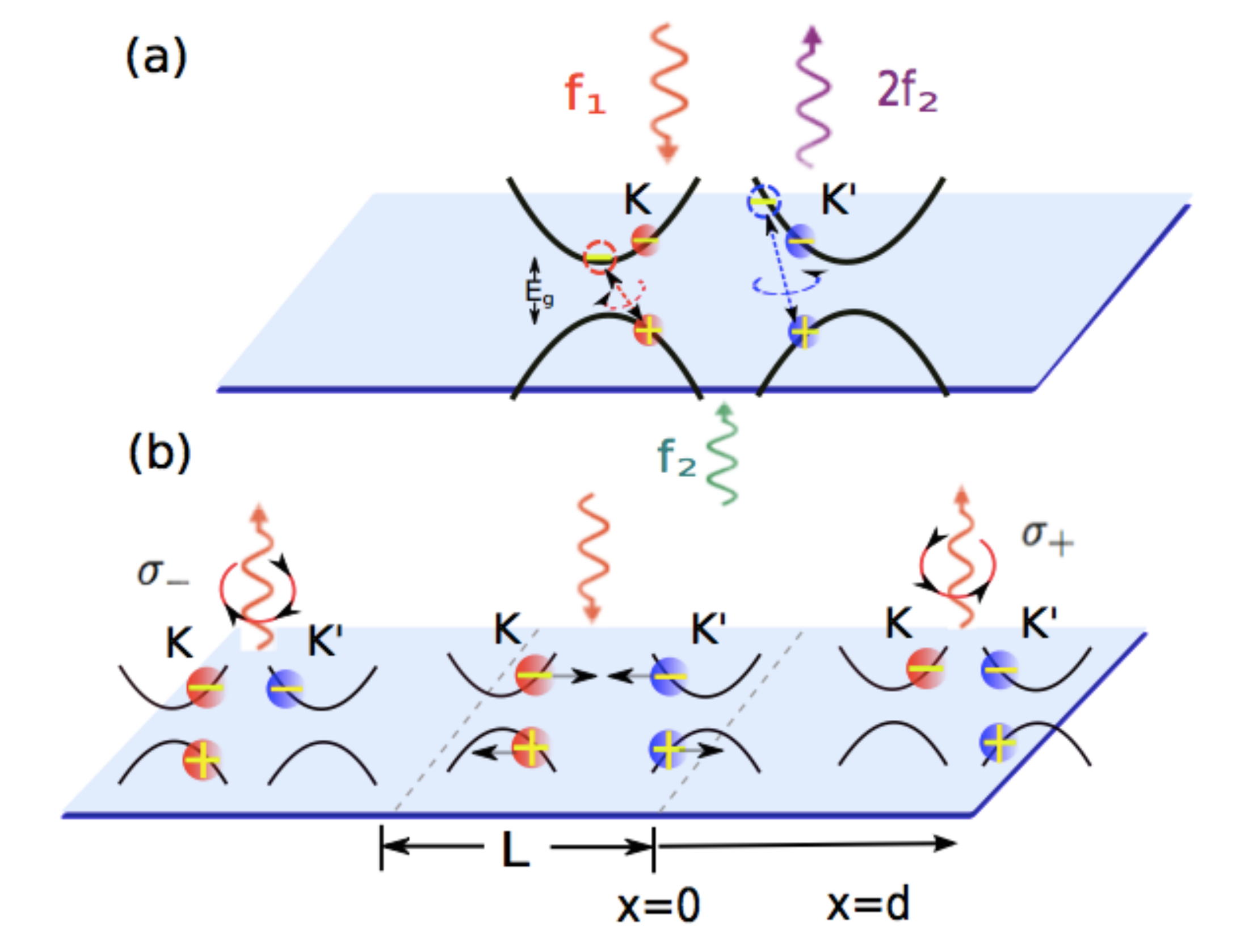}
\caption{(color online). Detection of pure valley current in monolayer MoS$_2$. (a) The second-harmonic generation from the pump-probe experiment. Red (green) line denotes pump (probe) light, and purple line refers to the second-harmonic generation. $f_1 (f_2)$ is the frequency of the pump (probe) light. (b) Photoluminescence (PL) behavior over the electron-hole recombination length $d$.  Red (blue) ball labels carriers from valley $K (K')$ and '-'('+') denotes electrons (holes).}
\label{fig:detection}
\end{figure}

An alternative proposal, unique to monolayer TMD, is to investigate the photoluminescence (PL) helicity as shown in Fig. \ref{fig:detection}(b).
Suppose a linearly polarized light illuminates the central region, generating a steady valley current.
As the valley carriers move outside of the central region, they will start the diffusion process described by
\begin{equation}
D\nabla^2\delta\mu(x)-\delta\mu(x)/\tau_{inter} = 0 \;,
\end{equation}
where $D = v_F^2\tau/2$ is the diffusion constant---derived from the Fermi velocity $v_F$ and the momentum relaxation time $\tau$---and $\delta\mu(x)=\mu^K(x)-\mu^{K'}(x)$ is
the chemical potential difference between the two valleys.  This equation describes both electrons and holes.
Consider the right region of the sample.  For a valley current with initial velocity $j^v=\pm(\sigma_{xx}/2e)\partial_x\delta\mu(x)$, where $+(-)$ corresponds to holes (electrons) and $\sigma_{xx}$ is the total longitudinal (Drude) conductivity, we obtain
\begin{align}
\delta\mu(x)=(\mp 2ej^v\ell_{inter}/\sigma_{xx})\exp(-x/\ell_{inter}) \;,
\end{align}
where $\ell_{inter} =\sqrt{D\tau_{inter}}$ is the valley diffusion length.
In monolayer TMD, we have $\tau^e_{inter} \ll \tau_{com} \ll \tau^h_{inter}$, where $\tau_{com}$ is the electron-hole recombination time.~\cite{mak2012}
Therefore, after a diffusion length $d\sim\sqrt{D\tau_{com}}$, holes will have a local chemical potential difference $\delta\mu_h(x) < 0$; meanwhile, electrons become almost equally populated in the two valleys and $\delta\mu_e(x) \simeq 0$.
Following the valley-contrasting circular dichroism,~\cite{xiao2012,cao2012,mak2012,zeng2012} this leads to a net $\sigma^+$ PL hecility. A similar argument can be applied to the left region of the sample, where a net $\sigma^-$ PL is expected.
Experimental results found~\cite{mak2012} that $\tau^h_{inter} \sim 1$ ns, which leads to $\ell^h_{inter} \sim 1$ $\mu$m; they further determined that $\tau_{com}=50$ ps, resulting in a PL helicity $(\mu^{K}-\mu^{K'})/(\mu^{K}+\mu^{K'})=0.8$.
Note that this phenomenon is intimately related to the spin-valley coupled bands, and hence absent in other multi-valleyed systems such as staggered monolayer or biased bilayer graphene.

\begin{table*}[t]
\caption{ Valley-resolved longitudinal (transverse) current $j_x^\eta(j_y^\eta)$ under oblique incidence for linear and circular photogalvanic and photon drag effect, with $\eta=K$ or $K'$. $\theta$ and $\phi$ are polarization and incident angle, respectively. $T_{xxxx}$, $T_{xxyy}$ and $T_{xxxy}$ are independent components of a rank-4 tensor. $E_0$ is the amplitude of the electric field.}
\label{tab:oblique}%
\begin{ruledtabular}
\begin{tabular}{ccc}
 &  Longitudinal ($j_x^\eta$)  & Transverse ($j_y^\eta$)   \\ \hline
Linear photogalvanic effect & $\chi_{xxx}^\eta(\cos^2\theta\cos^2\phi-\sin^2\theta)E_0^2$ & $-\chi_{xxx}^\eta\cos\phi\sin2\theta E_0^2$  \\
Circular photogalvanic effect & $-\chi_{xxx}^\eta\sin^2\phi E_0^2$ & 0 \\
Linear photon drag effect & $Q_x(T^\eta_{xxxx}\cos^2\theta\cos^2\phi+T^\eta_{xxyy}\sin^2\theta)E_0^2$ & $\frac{Q_x}{2}(T^\eta_{xxxx}-T^\eta_{xxyy})\cos\phi\sin2\theta E_0^2$ \\
Circular photon drag effect & $Q_x(T^\eta_{xxxx}\cos^2\phi+T^\eta_{xxyy}-2iT^\eta_{xxxy}\cos\phi)E_0^2$ & 0
\end{tabular}
\end{ruledtabular}
\end{table*}

\section{\label{sec:discussion}Discussion and conclusion}

So far we have considered the normal incidence case. 
For oblique incidence, the results are summarized in Table~\ref{tab:oblique}, where the general form of the valley-resolved current $j_x^\eta(j_y^\eta)$ along the longitudinal (transverse) direction is given.
$\eta$ refers to valleys $K$ and $K'$, and the $xz$ plane is the incident plane.
For a linearly polarized light, valley current is induced in both longitudinal and transverse directions; in contrast, only the longitudinal valley current is generated under circularly polarized light.
In both cases, charge current vanishes since $\chi_{xxx}^{K}+\chi_{xxx}^{K'}=0$ due to the mirror symmetry under $M_x$, leading to a pure valley current.
Apart from the PGE, there exists another photocurrent generating mechanism under the oblique incidence, namely the photon drag effect (PDE).~\cite{glazov2014}
In this case, photons transfer both momentum and angular momentum to carriers, and the current is described by $j_{\alpha}=T_{\alpha\beta\gamma\zeta}Q_{\beta}E_{\gamma}E_{\zeta}^*$, where $T$ is a rank-4 tensor and $\bm{Q}$ is the photon wavevector.
Similar to the PGE, PDE contributes to the valley current under both linearly and circularly polarized light.
However, $T_{xxxx}^K+T_{xxxx}^{K'}\neq0$, $T_{xxyy}^K+T_{xxyy}^{K'}\neq0$, indicating that the net charge current does not necessarily vanish in the system.
To distinguish these two mechanisms in experiments, one notices the fact that the response from PDE (PGE) is an odd (even) function under the reversal of incident direction $\bm Q\rightarrow-\bm{Q}$, by which the dominant mechanism can be identified.

The proposal of pure valley current generation can be generalized to other systems with appropriate symmetries, however, the magnitude of the effect and the detection scheme may vary among different systems. In our view, monolayer TMD has its advantage that the specific band and symmetries allow peculiar detection and observable signal.

To conclude, we have demonstrated that a linearly-polarized light can induce a pure valley current in monolayer TMD.  This mechanism originates from the $C_3$-symmetry rather than spin-orbit coupling.  Furthermore, we propose two realistic optical approaches to detect the pure valley current.
The effect of oblique incidence is also discussed.

\textit{Note added}.---Upon the completion of this work, we become aware of two recent papers, Ref.~\onlinecite{yu2014} and \onlinecite{sipe2014}, which also studied the nonlinear valley effect.

\section*{ Acknowledgement}

We are grateful to Xiaodong Xu, Wang Yao and Sanfeng Wu for stimulating discussions, and Matthew Daniels for a careful reading of the manuscript.  The theoretical part of this work was supported by DOE (No. DE-SC0012509) and the simulation part by AFOSR (No. FA9550-14-1-0277).


\begin{thebibliography}{51}%
\makeatletter
\providecommand \@ifxundefined [1]{%
 \@ifx{#1\undefined}
}%
\providecommand \@ifnum [1]{%
 \ifnum #1\expandafter \@firstoftwo
 \else \expandafter \@secondoftwo
 \fi
}%
\providecommand \@ifx [1]{%
 \ifx #1\expandafter \@firstoftwo
 \else \expandafter \@secondoftwo
 \fi
}%
\providecommand \natexlab [1]{#1}%
\providecommand \enquote  [1]{``#1''}%
\providecommand \bibnamefont  [1]{#1}%
\providecommand \bibfnamefont [1]{#1}%
\providecommand \citenamefont [1]{#1}%
\providecommand \href@noop [0]{\@secondoftwo}%
\providecommand \href [0]{\begingroup \@sanitize@url \@href}%
\providecommand \@href[1]{\@@startlink{#1}\@@href}%
\providecommand \@@href[1]{\endgroup#1\@@endlink}%
\providecommand \@sanitize@url [0]{\catcode `\\12\catcode `\$12\catcode
  `\&12\catcode `\#12\catcode `\^12\catcode `\_12\catcode `\%12\relax}%
\providecommand \@@startlink[1]{}%
\providecommand \@@endlink[0]{}%
\providecommand \url  [0]{\begingroup\@sanitize@url \@url }%
\providecommand \@url [1]{\endgroup\@href {#1}{\urlprefix }}%
\providecommand \urlprefix  [0]{URL }%
\providecommand \Eprint [0]{\href }%
\providecommand \doibase [0]{http://dx.doi.org/}%
\providecommand \selectlanguage [0]{\@gobble}%
\providecommand \bibinfo  [0]{\@secondoftwo}%
\providecommand \bibfield  [0]{\@secondoftwo}%
\providecommand \translation [1]{[#1]}%
\providecommand \BibitemOpen [0]{}%
\providecommand \bibitemStop [0]{}%
\providecommand \bibitemNoStop [0]{.\EOS\space}%
\providecommand \EOS [0]{\spacefactor3000\relax}%
\providecommand \BibitemShut  [1]{\csname bibitem#1\endcsname}%
\let\auto@bib@innerbib\@empty
\bibitem [{\citenamefont {Gunawan}\ \emph {et~al.}(2006)\citenamefont
  {Gunawan}, \citenamefont {Shkolnikov}, \citenamefont {Vakili}, \citenamefont
  {Gokmen}, \citenamefont {De~Poortere},\ and\ \citenamefont
  {Shayegan}}]{gunawan2006}%
  \BibitemOpen
  \bibfield  {author} {\bibinfo {author} {\bibfnamefont {O.}~\bibnamefont
  {Gunawan}}, \bibinfo {author} {\bibfnamefont {Y.~P.}\ \bibnamefont
  {Shkolnikov}}, \bibinfo {author} {\bibfnamefont {K.}~\bibnamefont {Vakili}},
  \bibinfo {author} {\bibfnamefont {T.}~\bibnamefont {Gokmen}}, \bibinfo
  {author} {\bibfnamefont {E.~P.}\ \bibnamefont {De~Poortere}}, \ and\ \bibinfo
  {author} {\bibfnamefont {M.}~\bibnamefont {Shayegan}},\ }\href {\doibase
  10.1103/PhysRevLett.97.186404} {\bibfield  {journal} {\bibinfo  {journal}
  {Phys. Rev. Lett.}\ }\textbf {\bibinfo {volume} {97}},\ \bibinfo {pages}
  {186404} (\bibinfo {year} {2006})}\BibitemShut {NoStop}%
\bibitem [{\citenamefont {Rycerz}\ \emph {et~al.}(2007)\citenamefont {Rycerz},
  \citenamefont {Tworzydlo},\ and\ \citenamefont {Beenakker}}]{rycerz2007}%
  \BibitemOpen
  \bibfield  {author} {\bibinfo {author} {\bibfnamefont {A.}~\bibnamefont
  {Rycerz}}, \bibinfo {author} {\bibfnamefont {J.}~\bibnamefont {Tworzydlo}}, \
  and\ \bibinfo {author} {\bibfnamefont {C.~W.~J.}\ \bibnamefont {Beenakker}},\
  }\href {\doibase 10.1038/nphys547} {\bibfield  {journal} {\bibinfo  {journal}
  {Nature Phys.}\ }\textbf {\bibinfo {volume} {3}},\ \bibinfo {pages} {172}
  (\bibinfo {year} {2007})}\BibitemShut {NoStop}%
\bibitem [{\citenamefont {Xiao}\ \emph {et~al.}(2007)\citenamefont {Xiao},
  \citenamefont {Yao},\ and\ \citenamefont {Niu}}]{xiao2007}%
  \BibitemOpen
  \bibfield  {author} {\bibinfo {author} {\bibfnamefont {D.}~\bibnamefont
  {Xiao}}, \bibinfo {author} {\bibfnamefont {W.}~\bibnamefont {Yao}}, \ and\
  \bibinfo {author} {\bibfnamefont {Q.}~\bibnamefont {Niu}},\ }\href {\doibase
  10.1103/PhysRevLett.99.236809} {\bibfield  {journal} {\bibinfo  {journal}
  {Phys. Rev. Lett.}\ }\textbf {\bibinfo {volume} {99}},\ \bibinfo {pages}
  {236809} (\bibinfo {year} {2007})}\BibitemShut {NoStop}%
\bibitem [{\citenamefont {Yao}\ \emph {et~al.}(2008)\citenamefont {Yao},
  \citenamefont {Xiao},\ and\ \citenamefont {Niu}}]{yao2008}%
  \BibitemOpen
  \bibfield  {author} {\bibinfo {author} {\bibfnamefont {W.}~\bibnamefont
  {Yao}}, \bibinfo {author} {\bibfnamefont {D.}~\bibnamefont {Xiao}}, \ and\
  \bibinfo {author} {\bibfnamefont {Q.}~\bibnamefont {Niu}},\ }\href {\doibase
  10.1103/PhysRevB.77.235406} {\bibfield  {journal} {\bibinfo  {journal} {Phys.
  Rev. B}\ }\textbf {\bibinfo {volume} {77}},\ \bibinfo {pages} {235406}
  (\bibinfo {year} {2008})}\BibitemShut {NoStop}%
\bibitem [{\citenamefont {Gunlycke}\ and\ \citenamefont
  {White}(2011)}]{gunlycke2011}%
  \BibitemOpen
  \bibfield  {author} {\bibinfo {author} {\bibfnamefont {D.}~\bibnamefont
  {Gunlycke}}\ and\ \bibinfo {author} {\bibfnamefont {C.~T.}\ \bibnamefont
  {White}},\ }\href {\doibase 10.1103/PhysRevLett.106.136806} {\bibfield
  {journal} {\bibinfo  {journal} {Phys. Rev. Lett.}\ }\textbf {\bibinfo
  {volume} {106}},\ \bibinfo {pages} {136806} (\bibinfo {year}
  {2011})}\BibitemShut {NoStop}%
\bibitem [{\citenamefont {Zhu}\ \emph {et~al.}(2012)\citenamefont {Zhu},
  \citenamefont {Collaudin}, \citenamefont {Fauque}, \citenamefont {Kang},\
  and\ \citenamefont {Behnia}}]{zhu2012}%
  \BibitemOpen
  \bibfield  {author} {\bibinfo {author} {\bibfnamefont {Z.}~\bibnamefont
  {Zhu}}, \bibinfo {author} {\bibfnamefont {A.}~\bibnamefont {Collaudin}},
  \bibinfo {author} {\bibfnamefont {B.}~\bibnamefont {Fauque}}, \bibinfo
  {author} {\bibfnamefont {W.}~\bibnamefont {Kang}}, \ and\ \bibinfo {author}
  {\bibfnamefont {K.}~\bibnamefont {Behnia}},\ }\href {\doibase
  10.1038/nphys2111} {\bibfield  {journal} {\bibinfo  {journal} {Nature Phys.}\
  }\textbf {\bibinfo {volume} {8}},\ \bibinfo {pages} {89} (\bibinfo {year}
  {2012})}\BibitemShut {NoStop}%
\bibitem [{\citenamefont {Jiang}\ \emph {et~al.}(2013)\citenamefont {Jiang},
  \citenamefont {Low}, \citenamefont {Chang}, \citenamefont {Katsnelson},\ and\
  \citenamefont {Guinea}}]{jiang2013}%
  \BibitemOpen
  \bibfield  {author} {\bibinfo {author} {\bibfnamefont {Y.}~\bibnamefont
  {Jiang}}, \bibinfo {author} {\bibfnamefont {T.}~\bibnamefont {Low}}, \bibinfo
  {author} {\bibfnamefont {K.}~\bibnamefont {Chang}}, \bibinfo {author}
  {\bibfnamefont {M.~I.}\ \bibnamefont {Katsnelson}}, \ and\ \bibinfo {author}
  {\bibfnamefont {F.}~\bibnamefont {Guinea}},\ }\href {\doibase
  10.1103/PhysRevLett.110.046601} {\bibfield  {journal} {\bibinfo  {journal}
  {Phys. Rev. Lett.}\ }\textbf {\bibinfo {volume} {110}},\ \bibinfo {pages}
  {046601} (\bibinfo {year} {2013})}\BibitemShut {NoStop}%
\bibitem [{\citenamefont {Isberg}\ \emph {et~al.}(2013)\citenamefont {Isberg},
  \citenamefont {Gabrysch}, \citenamefont {Hammersberg}, \citenamefont {Majdi},
  \citenamefont {Kovi},\ and\ \citenamefont {Twitchen}}]{isberg2013}%
  \BibitemOpen
  \bibfield  {author} {\bibinfo {author} {\bibfnamefont {J.}~\bibnamefont
  {Isberg}}, \bibinfo {author} {\bibfnamefont {M.}~\bibnamefont {Gabrysch}},
  \bibinfo {author} {\bibfnamefont {J.}~\bibnamefont {Hammersberg}}, \bibinfo
  {author} {\bibfnamefont {S.}~\bibnamefont {Majdi}}, \bibinfo {author}
  {\bibfnamefont {K.~K.}\ \bibnamefont {Kovi}}, \ and\ \bibinfo {author}
  {\bibfnamefont {D.~J.}\ \bibnamefont {Twitchen}},\ }\href {\doibase
  10.1038/nmat3694} {\bibfield  {journal} {\bibinfo  {journal} {Nature Mater.}\
  }\textbf {\bibinfo {volume} {12}},\ \bibinfo {pages} {760} (\bibinfo {year}
  {2013})}\BibitemShut {NoStop}%
\bibitem [{\citenamefont {Xu}\ \emph {et~al.}(2014)\citenamefont {Xu},
  \citenamefont {Yao}, \citenamefont {Xiao},\ and\ \citenamefont
  {Heinz}}]{natphys}%
  \BibitemOpen
  \bibfield  {author} {\bibinfo {author} {\bibfnamefont {X.}~\bibnamefont
  {Xu}}, \bibinfo {author} {\bibfnamefont {W.}~\bibnamefont {Yao}}, \bibinfo
  {author} {\bibfnamefont {D.}~\bibnamefont {Xiao}}, \ and\ \bibinfo {author}
  {\bibfnamefont {T.~F.}\ \bibnamefont {Heinz}},\ }\href@noop {} {\bibfield
  {journal} {\bibinfo  {journal} {Nature Phys.}\ }\textbf {\bibinfo {volume}
  {10}},\ \bibinfo {pages} {343} (\bibinfo {year} {2014})}\BibitemShut
  {NoStop}%
\bibitem [{\citenamefont {Xiao}\ \emph {et~al.}(2012)\citenamefont {Xiao},
  \citenamefont {Liu}, \citenamefont {Feng}, \citenamefont {Xu},\ and\
  \citenamefont {Yao}}]{xiao2012}%
  \BibitemOpen
  \bibfield  {author} {\bibinfo {author} {\bibfnamefont {D.}~\bibnamefont
  {Xiao}}, \bibinfo {author} {\bibfnamefont {G.-B.}\ \bibnamefont {Liu}},
  \bibinfo {author} {\bibfnamefont {W.}~\bibnamefont {Feng}}, \bibinfo {author}
  {\bibfnamefont {X.}~\bibnamefont {Xu}}, \ and\ \bibinfo {author}
  {\bibfnamefont {W.}~\bibnamefont {Yao}},\ }\href {\doibase
  10.1103/PhysRevLett.108.196802} {\bibfield  {journal} {\bibinfo  {journal}
  {Phys. Rev. Lett.}\ }\textbf {\bibinfo {volume} {108}},\ \bibinfo {pages}
  {196802} (\bibinfo {year} {2012})}\BibitemShut {NoStop}%
\bibitem [{\citenamefont {Cao}\ \emph {et~al.}(2012)\citenamefont {Cao},
  \citenamefont {Wang}, \citenamefont {Han}, \citenamefont {Ye}, \citenamefont
  {Zhu}, \citenamefont {Shi}, \citenamefont {Niu}, \citenamefont {Tan},
  \citenamefont {Wang}, \citenamefont {Liu},\ and\ \citenamefont
  {Feng}}]{cao2012}%
  \BibitemOpen
  \bibfield  {author} {\bibinfo {author} {\bibfnamefont {T.}~\bibnamefont
  {Cao}}, \bibinfo {author} {\bibfnamefont {G.}~\bibnamefont {Wang}}, \bibinfo
  {author} {\bibfnamefont {W.}~\bibnamefont {Han}}, \bibinfo {author}
  {\bibfnamefont {H.}~\bibnamefont {Ye}}, \bibinfo {author} {\bibfnamefont
  {C.}~\bibnamefont {Zhu}}, \bibinfo {author} {\bibfnamefont {J.}~\bibnamefont
  {Shi}}, \bibinfo {author} {\bibfnamefont {Q.}~\bibnamefont {Niu}}, \bibinfo
  {author} {\bibfnamefont {P.}~\bibnamefont {Tan}}, \bibinfo {author}
  {\bibfnamefont {E.}~\bibnamefont {Wang}}, \bibinfo {author} {\bibfnamefont
  {B.}~\bibnamefont {Liu}}, \ and\ \bibinfo {author} {\bibfnamefont
  {J.}~\bibnamefont {Feng}},\ }\href@noop {} {\bibfield  {journal} {\bibinfo
  {journal} {Nat. Commun.}\ }\textbf {\bibinfo {volume} {3}},\ \bibinfo {pages}
  {887} (\bibinfo {year} {2012})}\BibitemShut {NoStop}%
\bibitem [{\citenamefont {Zeng}\ \emph {et~al.}(2012)\citenamefont {Zeng},
  \citenamefont {Dai}, \citenamefont {Yao}, \citenamefont {Xiao},\ and\
  \citenamefont {Cui}}]{zeng2012}%
  \BibitemOpen
  \bibfield  {author} {\bibinfo {author} {\bibfnamefont {H.}~\bibnamefont
  {Zeng}}, \bibinfo {author} {\bibfnamefont {J.}~\bibnamefont {Dai}}, \bibinfo
  {author} {\bibfnamefont {W.}~\bibnamefont {Yao}}, \bibinfo {author}
  {\bibfnamefont {D.}~\bibnamefont {Xiao}}, \ and\ \bibinfo {author}
  {\bibfnamefont {X.}~\bibnamefont {Cui}},\ }\href@noop {} {\bibfield
  {journal} {\bibinfo  {journal} {Nat. Nanotech.}\ }\textbf {\bibinfo {volume}
  {7}},\ \bibinfo {pages} {490} (\bibinfo {year} {2012})}\BibitemShut {NoStop}%
\bibitem [{\citenamefont {Mak}\ \emph {et~al.}(2012)\citenamefont {Mak},
  \citenamefont {He}, \citenamefont {Shan},\ and\ \citenamefont
  {Heinz}}]{mak2012}%
  \BibitemOpen
  \bibfield  {author} {\bibinfo {author} {\bibfnamefont {K.~F.}\ \bibnamefont
  {Mak}}, \bibinfo {author} {\bibfnamefont {K.}~\bibnamefont {He}}, \bibinfo
  {author} {\bibfnamefont {J.}~\bibnamefont {Shan}}, \ and\ \bibinfo {author}
  {\bibfnamefont {T.~F.}\ \bibnamefont {Heinz}},\ }\href@noop {} {\bibfield
  {journal} {\bibinfo  {journal} {Nat. Nanotech.}\ }\textbf {\bibinfo {volume}
  {7}},\ \bibinfo {pages} {494} (\bibinfo {year} {2012})}\BibitemShut {NoStop}%
\bibitem [{\citenamefont {Mak}\ \emph {et~al.}(2014)\citenamefont {Mak},
  \citenamefont {McGill}, \citenamefont {Park},\ and\ \citenamefont
  {McEuen}}]{mak2014}%
  \BibitemOpen
  \bibfield  {author} {\bibinfo {author} {\bibfnamefont {K.~F.}\ \bibnamefont
  {Mak}}, \bibinfo {author} {\bibfnamefont {K.~L.}\ \bibnamefont {McGill}},
  \bibinfo {author} {\bibfnamefont {J.}~\bibnamefont {Park}}, \ and\ \bibinfo
  {author} {\bibfnamefont {P.~L.}\ \bibnamefont {McEuen}},\ }\href@noop {}
  {\bibfield  {journal} {\bibinfo  {journal} {Science}\ }\textbf {\bibinfo
  {volume} {344}},\ \bibinfo {pages} {1489} (\bibinfo {year}
  {2014})}\BibitemShut {NoStop}%
\bibitem [{\citenamefont {Yuan}\ \emph {et~al.}(2014)\citenamefont {Yuan},
  \citenamefont {Wang}, \citenamefont {Lian}, \citenamefont {Zhang},
  \citenamefont {Fang}, \citenamefont {Shen}, \citenamefont {Xu}, \citenamefont
  {Xu}, \citenamefont {Zhang}, \citenamefont {Hwang},\ and\ \citenamefont
  {Cui}}]{yuan2014}%
  \BibitemOpen
  \bibfield  {author} {\bibinfo {author} {\bibfnamefont {H.}~\bibnamefont
  {Yuan}}, \bibinfo {author} {\bibfnamefont {X.}~\bibnamefont {Wang}}, \bibinfo
  {author} {\bibfnamefont {B.}~\bibnamefont {Lian}}, \bibinfo {author}
  {\bibfnamefont {H.}~\bibnamefont {Zhang}}, \bibinfo {author} {\bibfnamefont
  {X.}~\bibnamefont {Fang}}, \bibinfo {author} {\bibfnamefont {B.}~\bibnamefont
  {Shen}}, \bibinfo {author} {\bibfnamefont {G.}~\bibnamefont {Xu}}, \bibinfo
  {author} {\bibfnamefont {Y.}~\bibnamefont {Xu}}, \bibinfo {author}
  {\bibfnamefont {S.-C.}\ \bibnamefont {Zhang}}, \bibinfo {author}
  {\bibfnamefont {H.~Y.}\ \bibnamefont {Hwang}}, \ and\ \bibinfo {author}
  {\bibfnamefont {Y.}~\bibnamefont {Cui}},\ }\href@noop {} {\bibfield
  {journal} {\bibinfo  {journal} {Nat. Nanotech.}\ }\textbf {\bibinfo {volume}
  {9}},\ \bibinfo {pages} {851} (\bibinfo {year} {2014})}\BibitemShut {NoStop}%
\bibitem [{\citenamefont {Zhang}\ \emph
  {et~al.}(2014{\natexlab{a}})\citenamefont {Zhang}, \citenamefont {Oka},
  \citenamefont {Suzuki}, \citenamefont {Ye},\ and\ \citenamefont
  {Iwasa}}]{zhang2014a}%
  \BibitemOpen
  \bibfield  {author} {\bibinfo {author} {\bibfnamefont {Y.~J.}\ \bibnamefont
  {Zhang}}, \bibinfo {author} {\bibfnamefont {T.}~\bibnamefont {Oka}}, \bibinfo
  {author} {\bibfnamefont {R.}~\bibnamefont {Suzuki}}, \bibinfo {author}
  {\bibfnamefont {J.~T.}\ \bibnamefont {Ye}}, \ and\ \bibinfo {author}
  {\bibfnamefont {Y.}~\bibnamefont {Iwasa}},\ }\href {\doibase
  10.1126/science.1251329} {\bibfield  {journal} {\bibinfo  {journal}
  {Science}\ }\textbf {\bibinfo {volume} {344}},\ \bibinfo {pages} {725}
  (\bibinfo {year} {2014}{\natexlab{a}})}\BibitemShut {NoStop}%
\bibitem [{\citenamefont {Jedema}\ \emph {et~al.}(2001)\citenamefont {Jedema},
  \citenamefont {Filip},\ and\ \citenamefont {van Wees}}]{jedema2001}%
  \BibitemOpen
  \bibfield  {author} {\bibinfo {author} {\bibfnamefont {F.~J.}\ \bibnamefont
  {Jedema}}, \bibinfo {author} {\bibfnamefont {A.~T.}\ \bibnamefont {Filip}}, \
  and\ \bibinfo {author} {\bibfnamefont {B.~J.}\ \bibnamefont {van Wees}},\
  }\href@noop {} {\bibfield  {journal} {\bibinfo  {journal} {Science}\ }\textbf
  {\bibinfo {volume} {410}},\ \bibinfo {pages} {345} (\bibinfo {year}
  {2001})}\BibitemShut {NoStop}%
\bibitem [{\citenamefont {Jedema}\ \emph {et~al.}(2003)\citenamefont {Jedema},
  \citenamefont {Nijboer}, \citenamefont {Filip},\ and\ \citenamefont {van
  Wees}}]{jedema2003}%
  \BibitemOpen
  \bibfield  {author} {\bibinfo {author} {\bibfnamefont {F.~J.}\ \bibnamefont
  {Jedema}}, \bibinfo {author} {\bibfnamefont {M.~S.}\ \bibnamefont {Nijboer}},
  \bibinfo {author} {\bibfnamefont {A.~T.}\ \bibnamefont {Filip}}, \ and\
  \bibinfo {author} {\bibfnamefont {B.~J.}\ \bibnamefont {van Wees}},\
  }\href@noop {} {\bibfield  {journal} {\bibinfo  {journal} {Phys. Rev. B}\
  }\textbf {\bibinfo {volume} {67}},\ \bibinfo {pages} {085319} (\bibinfo
  {year} {2003})}\BibitemShut {NoStop}%
\bibitem [{\citenamefont {Yang}\ \emph {et~al.}(2008)\citenamefont {Yang},
  \citenamefont {Kimura},\ and\ \citenamefont {Otani}}]{yang2008}%
  \BibitemOpen
  \bibfield  {author} {\bibinfo {author} {\bibfnamefont {T.}~\bibnamefont
  {Yang}}, \bibinfo {author} {\bibfnamefont {T.}~\bibnamefont {Kimura}}, \ and\
  \bibinfo {author} {\bibfnamefont {Y.}~\bibnamefont {Otani}},\ }\href@noop {}
  {\bibfield  {journal} {\bibinfo  {journal} {Nature Phys.}\ }\textbf {\bibinfo
  {volume} {4}},\ \bibinfo {pages} {851} (\bibinfo {year} {2008})}\BibitemShut
  {NoStop}%
\bibitem [{\citenamefont {Murakami}\ \emph {et~al.}(2003)\citenamefont
  {Murakami}, \citenamefont {Nagaosa},\ and\ \citenamefont
  {Zhang}}]{murakami2003}%
  \BibitemOpen
  \bibfield  {author} {\bibinfo {author} {\bibfnamefont {S.}~\bibnamefont
  {Murakami}}, \bibinfo {author} {\bibfnamefont {N.}~\bibnamefont {Nagaosa}}, \
  and\ \bibinfo {author} {\bibfnamefont {S.-C.}\ \bibnamefont {Zhang}},\
  }\href@noop {} {\bibfield  {journal} {\bibinfo  {journal} {Science}\ }\textbf
  {\bibinfo {volume} {301}},\ \bibinfo {pages} {1348} (\bibinfo {year}
  {2003})}\BibitemShut {NoStop}%
\bibitem [{\citenamefont {Bhat}\ \emph {et~al.}(2005)\citenamefont {Bhat},
  \citenamefont {Nastos}, \citenamefont {Najmaie},\ and\ \citenamefont
  {Sipe}}]{bhat2005}%
  \BibitemOpen
  \bibfield  {author} {\bibinfo {author} {\bibfnamefont {R.~D.~R.}\
  \bibnamefont {Bhat}}, \bibinfo {author} {\bibfnamefont {F.}~\bibnamefont
  {Nastos}}, \bibinfo {author} {\bibfnamefont {A.}~\bibnamefont {Najmaie}}, \
  and\ \bibinfo {author} {\bibfnamefont {J.~E.}\ \bibnamefont {Sipe}},\ }\href
  {\doibase 10.1103/PhysRevLett.94.096603} {\bibfield  {journal} {\bibinfo
  {journal} {Phys. Rev. Lett.}\ }\textbf {\bibinfo {volume} {94}},\ \bibinfo
  {pages} {096603} (\bibinfo {year} {2005})}\BibitemShut {NoStop}%
\bibitem [{\citenamefont {Kraut}\ and\ \citenamefont {von
  Baltz}(1979)}]{kraut1979}%
  \BibitemOpen
  \bibfield  {author} {\bibinfo {author} {\bibfnamefont {W.}~\bibnamefont
  {Kraut}}\ and\ \bibinfo {author} {\bibfnamefont {R.}~\bibnamefont {von
  Baltz}},\ }\href {\doibase 10.1103/PhysRevB.19.1548} {\bibfield  {journal}
  {\bibinfo  {journal} {Phys. Rev. B}\ }\textbf {\bibinfo {volume} {19}},\
  \bibinfo {pages} {1548} (\bibinfo {year} {1979})}\BibitemShut {NoStop}%
\bibitem [{boy()}]{boyd2008}%
  \BibitemOpen
  \href@noop {} {}\bibinfo {note} {R. Boyd, \emph{Nonlinear Optics}, 3rd ed.
  (Academic Press, Burlington, MA, 2008).}\BibitemShut {Stop}%
\bibitem [{\citenamefont {Hosur}(2011)}]{hosur2011}%
  \BibitemOpen
  \bibfield  {author} {\bibinfo {author} {\bibfnamefont {P.}~\bibnamefont
  {Hosur}},\ }\href {\doibase 10.1103/PhysRevB.83.035309} {\bibfield  {journal}
  {\bibinfo  {journal} {Phys. Rev. B}\ }\textbf {\bibinfo {volume} {83}},\
  \bibinfo {pages} {035309} (\bibinfo {year} {2011})}\BibitemShut {NoStop}%
\bibitem [{Note1()}]{Note1}%
  \BibitemOpen
  \bibinfo {note} {In low-quality sample or system with small optical
  transition gap, the linear photogalvanic effect is non-negligible and valley
  current becomes partially polarized. This may explain the observed
  photocurrent in Ref. \protect \rev@citealpnum {yuan2014}.}\BibitemShut
  {Stop}%
\bibitem [{stu()}]{sturman1992}%
  \BibitemOpen
  \href@noop {} {}\bibinfo {note} {B. I. Sturman and V. M. Fridkin, \emph{The
  Photovoltaic and Photorefractive Effects in Noncentrosymmetric Materials}
  (Gordon and Breach, Philadelphia, 1992).}\BibitemShut {Stop}%
\bibitem [{\citenamefont {Sipe}\ and\ \citenamefont
  {Shkrebtii}(2000)}]{sipe2000}%
  \BibitemOpen
  \bibfield  {author} {\bibinfo {author} {\bibfnamefont {J.~E.}\ \bibnamefont
  {Sipe}}\ and\ \bibinfo {author} {\bibfnamefont {A.~I.}\ \bibnamefont
  {Shkrebtii}},\ }\href {\doibase 10.1103/PhysRevB.61.5337} {\bibfield
  {journal} {\bibinfo  {journal} {Phys. Rev. B}\ }\textbf {\bibinfo {volume}
  {61}},\ \bibinfo {pages} {5337} (\bibinfo {year} {2000})}\BibitemShut
  {NoStop}%
\bibitem [{\citenamefont {Lu}\ \emph {et~al.}(2013)\citenamefont {Lu},
  \citenamefont {Yao}, \citenamefont {Xiao},\ and\ \citenamefont
  {Shen}}]{lu2013}%
  \BibitemOpen
  \bibfield  {author} {\bibinfo {author} {\bibfnamefont {H.-Z.}\ \bibnamefont
  {Lu}}, \bibinfo {author} {\bibfnamefont {W.}~\bibnamefont {Yao}}, \bibinfo
  {author} {\bibfnamefont {D.}~\bibnamefont {Xiao}}, \ and\ \bibinfo {author}
  {\bibfnamefont {S.-Q.}\ \bibnamefont {Shen}},\ }\href {\doibase
  10.1103/PhysRevLett.110.016806} {\bibfield  {journal} {\bibinfo  {journal}
  {Phys. Rev. Lett.}\ }\textbf {\bibinfo {volume} {110}},\ \bibinfo {pages}
  {016806} (\bibinfo {year} {2013})}\BibitemShut {NoStop}%
\bibitem [{\citenamefont {Song}\ and\ \citenamefont {Dery}(2013)}]{song2013}%
  \BibitemOpen
  \bibfield  {author} {\bibinfo {author} {\bibfnamefont {Y.}~\bibnamefont
  {Song}}\ and\ \bibinfo {author} {\bibfnamefont {H.}~\bibnamefont {Dery}},\
  }\href {\doibase 10.1103/PhysRevLett.111.026601} {\bibfield  {journal}
  {\bibinfo  {journal} {Phys. Rev. Lett.}\ }\textbf {\bibinfo {volume} {111}},\
  \bibinfo {pages} {026601} (\bibinfo {year} {2013})}\BibitemShut {NoStop}%
\bibitem [{\citenamefont {Shan}\ \emph {et~al.}(2013)\citenamefont {Shan},
  \citenamefont {Lu},\ and\ \citenamefont {Xiao}}]{shan2013}%
  \BibitemOpen
  \bibfield  {author} {\bibinfo {author} {\bibfnamefont {W.-Y.}\ \bibnamefont
  {Shan}}, \bibinfo {author} {\bibfnamefont {H.-Z.}\ \bibnamefont {Lu}}, \ and\
  \bibinfo {author} {\bibfnamefont {D.}~\bibnamefont {Xiao}},\ }\href {\doibase
  10.1103/PhysRevB.88.125301} {\bibfield  {journal} {\bibinfo  {journal} {Phys.
  Rev. B}\ }\textbf {\bibinfo {volume} {88}},\ \bibinfo {pages} {125301}
  (\bibinfo {year} {2013})}\BibitemShut {NoStop}%
\bibitem [{\citenamefont {Ochoa}\ and\ \citenamefont
  {Rold\'an}(2013)}]{ochoa2013}%
  \BibitemOpen
  \bibfield  {author} {\bibinfo {author} {\bibfnamefont {H.}~\bibnamefont
  {Ochoa}}\ and\ \bibinfo {author} {\bibfnamefont {R.}~\bibnamefont
  {Rold\'an}},\ }\href {\doibase 10.1103/PhysRevB.87.245421} {\bibfield
  {journal} {\bibinfo  {journal} {Phys. Rev. B}\ }\textbf {\bibinfo {volume}
  {87}},\ \bibinfo {pages} {245421} (\bibinfo {year} {2013})}\BibitemShut
  {NoStop}%
\bibitem [{\citenamefont {Liu}\ \emph {et~al.}(2013)\citenamefont {Liu},
  \citenamefont {Shan}, \citenamefont {Yao}, \citenamefont {Yao},\ and\
  \citenamefont {Xiao}}]{liu2013}%
  \BibitemOpen
  \bibfield  {author} {\bibinfo {author} {\bibfnamefont {G.-B.}\ \bibnamefont
  {Liu}}, \bibinfo {author} {\bibfnamefont {W.-Y.}\ \bibnamefont {Shan}},
  \bibinfo {author} {\bibfnamefont {Y.}~\bibnamefont {Yao}}, \bibinfo {author}
  {\bibfnamefont {W.}~\bibnamefont {Yao}}, \ and\ \bibinfo {author}
  {\bibfnamefont {D.}~\bibnamefont {Xiao}},\ }\href {\doibase
  10.1103/PhysRevB.88.085433} {\bibfield  {journal} {\bibinfo  {journal} {Phys.
  Rev. B}\ }\textbf {\bibinfo {volume} {88}},\ \bibinfo {pages} {085433}
  (\bibinfo {year} {2013})}\BibitemShut {NoStop}%
\bibitem [{Note2()}]{Note2}%
  \BibitemOpen
  \bibinfo {note} {Let us consider a laser beam with power $I=100$ $\mu $W and
  the light spot radius $r=1$ $\mu $m. We assume an absorption coefficient $\xi
  =2\%$.~\cite {mak2013,liu2014} By making use of the formula $\xi P=\protect
  \frac {1}{2}\epsilon _0\protect \sqrt {\epsilon _r}cE^2$ with vacuum
  dielectric constant $\epsilon _0$ and in-plane relative permittivity
  $\epsilon _r=2.8$,~\cite {cheiwchanchamnangij2012} we obtain the electric
  field $E = 3.01\times 10^4$ V/m. In addition, the mobility of monolayer
  MoS$_2$ is $\mu =200$ cm$^2$V$^{-1}$s$^{-1}$,~\cite {radisavljevic2011} which
  gives the momentum relaxation time $\tau =\mu m/|e|=55$ fs with $m$ being the
  effective mass.}\BibitemShut {Stop}%
\bibitem [{\citenamefont {Ganichev}\ and\ \citenamefont
  {Prettl}(2003)}]{ganichev2003}%
  \BibitemOpen
  \bibfield  {author} {\bibinfo {author} {\bibfnamefont {S.~D.}\ \bibnamefont
  {Ganichev}}\ and\ \bibinfo {author} {\bibfnamefont {W.}~\bibnamefont
  {Prettl}},\ }\href@noop {} {\bibfield  {journal} {\bibinfo  {journal} {J.
  Phys. Condens. Matter}\ }\textbf {\bibinfo {volume} {15}},\ \bibinfo {pages}
  {R935} (\bibinfo {year} {2003})}\BibitemShut {NoStop}%
\bibitem [{\citenamefont {Yang}\ and\ \citenamefont {Liu}(2013)}]{yang2013}%
  \BibitemOpen
  \bibfield  {author} {\bibinfo {author} {\bibfnamefont {F.}~\bibnamefont
  {Yang}}\ and\ \bibinfo {author} {\bibfnamefont {R.-B.}\ \bibnamefont {Liu}},\
  }\href@noop {} {\bibfield  {journal} {\bibinfo  {journal} {New J. Phys.}\
  }\textbf {\bibinfo {volume} {15}},\ \bibinfo {pages} {115005} (\bibinfo
  {year} {2013})}\BibitemShut {NoStop}%
\bibitem [{\citenamefont {Zhang}\ \emph
  {et~al.}(2014{\natexlab{b}})\citenamefont {Zhang}, \citenamefont {Johnson},
  \citenamefont {Hsu}, \citenamefont {Li},\ and\ \citenamefont
  {Shih}}]{zhang2014}%
  \BibitemOpen
  \bibfield  {author} {\bibinfo {author} {\bibfnamefont {C.}~\bibnamefont
  {Zhang}}, \bibinfo {author} {\bibfnamefont {A.}~\bibnamefont {Johnson}},
  \bibinfo {author} {\bibfnamefont {C.-L.}\ \bibnamefont {Hsu}}, \bibinfo
  {author} {\bibfnamefont {L.-J.}\ \bibnamefont {Li}}, \ and\ \bibinfo {author}
  {\bibfnamefont {C.-K.}\ \bibnamefont {Shih}},\ }\href@noop {} {\bibfield
  {journal} {\bibinfo  {journal} {Nano Lett.}\ }\textbf {\bibinfo {volume}
  {14}},\ \bibinfo {pages} {2443} (\bibinfo {year}
  {2014}{\natexlab{b}})}\BibitemShut {NoStop}%
\bibitem [{\citenamefont {Chernikov}\ \emph {et~al.}(2014)\citenamefont
  {Chernikov}, \citenamefont {Berkelbach}, \citenamefont {Hill}, \citenamefont
  {Rigosi}, \citenamefont {Li}, \citenamefont {Aslan}, \citenamefont
  {Reichman}, \citenamefont {Hybertsen},\ and\ \citenamefont
  {Heinz}}]{chernikov2014}%
  \BibitemOpen
  \bibfield  {author} {\bibinfo {author} {\bibfnamefont {A.}~\bibnamefont
  {Chernikov}}, \bibinfo {author} {\bibfnamefont {T.~C.}\ \bibnamefont
  {Berkelbach}}, \bibinfo {author} {\bibfnamefont {H.~M.}\ \bibnamefont
  {Hill}}, \bibinfo {author} {\bibfnamefont {A.}~\bibnamefont {Rigosi}},
  \bibinfo {author} {\bibfnamefont {Y.}~\bibnamefont {Li}}, \bibinfo {author}
  {\bibfnamefont {O.~B.}\ \bibnamefont {Aslan}}, \bibinfo {author}
  {\bibfnamefont {D.~R.}\ \bibnamefont {Reichman}}, \bibinfo {author}
  {\bibfnamefont {M.~S.}\ \bibnamefont {Hybertsen}}, \ and\ \bibinfo {author}
  {\bibfnamefont {T.~F.}\ \bibnamefont {Heinz}},\ }\href@noop {} {\bibfield
  {journal} {\bibinfo  {journal} {Phys. Rev. Lett.}\ }\textbf {\bibinfo
  {volume} {113}},\ \bibinfo {pages} {076802} (\bibinfo {year}
  {2014})}\BibitemShut {NoStop}%
\bibitem [{\citenamefont {Zhu}\ \emph {et~al.}(2014)\citenamefont {Zhu},
  \citenamefont {Chen},\ and\ \citenamefont {Cui}}]{zhu2014}%
  \BibitemOpen
  \bibfield  {author} {\bibinfo {author} {\bibfnamefont {B.}~\bibnamefont
  {Zhu}}, \bibinfo {author} {\bibfnamefont {X.}~\bibnamefont {Chen}}, \ and\
  \bibinfo {author} {\bibfnamefont {X.}~\bibnamefont {Cui}},\ }\href@noop {}
  {\bibfield  {journal} {\bibinfo  {journal} {arXiv:1403.5108}\ } (\bibinfo
  {year} {2014})}\BibitemShut {NoStop}%
\bibitem [{\citenamefont {Wang}\ \emph {et~al.}(2014)\citenamefont {Wang},
  \citenamefont {Marie}, \citenamefont {Gerber}, \citenamefont {Amand},
  \citenamefont {Lagarde}, \citenamefont {Bouet}, \citenamefont {Vidal},
  \citenamefont {Balocchi},\ and\ \citenamefont {Urbaszek}}]{wang2014}%
  \BibitemOpen
  \bibfield  {author} {\bibinfo {author} {\bibfnamefont {G.}~\bibnamefont
  {Wang}}, \bibinfo {author} {\bibfnamefont {X.}~\bibnamefont {Marie}},
  \bibinfo {author} {\bibfnamefont {I.}~\bibnamefont {Gerber}}, \bibinfo
  {author} {\bibfnamefont {T.}~\bibnamefont {Amand}}, \bibinfo {author}
  {\bibfnamefont {D.}~\bibnamefont {Lagarde}}, \bibinfo {author} {\bibfnamefont
  {L.}~\bibnamefont {Bouet}}, \bibinfo {author} {\bibfnamefont
  {M.}~\bibnamefont {Vidal}}, \bibinfo {author} {\bibfnamefont
  {A.}~\bibnamefont {Balocchi}}, \ and\ \bibinfo {author} {\bibfnamefont
  {B.}~\bibnamefont {Urbaszek}},\ }\href@noop {} {\bibfield  {journal}
  {\bibinfo  {journal} {arXiv:1404.0056}\ } (\bibinfo {year}
  {2014})}\BibitemShut {NoStop}%
\bibitem [{\citenamefont {He}\ \emph {et~al.}(2014)\citenamefont {He},
  \citenamefont {Kumar}, \citenamefont {Zhao}, \citenamefont {Wang},
  \citenamefont {Mak}, \citenamefont {Zhao},\ and\ \citenamefont
  {Shan}}]{he2014}%
  \BibitemOpen
  \bibfield  {author} {\bibinfo {author} {\bibfnamefont {K.}~\bibnamefont
  {He}}, \bibinfo {author} {\bibfnamefont {N.}~\bibnamefont {Kumar}}, \bibinfo
  {author} {\bibfnamefont {L.}~\bibnamefont {Zhao}}, \bibinfo {author}
  {\bibfnamefont {Z.}~\bibnamefont {Wang}}, \bibinfo {author} {\bibfnamefont
  {K.~F.}\ \bibnamefont {Mak}}, \bibinfo {author} {\bibfnamefont
  {H.}~\bibnamefont {Zhao}}, \ and\ \bibinfo {author} {\bibfnamefont
  {J.}~\bibnamefont {Shan}},\ }\href@noop {} {\bibfield  {journal} {\bibinfo
  {journal} {arXiv:1406.3095}\ } (\bibinfo {year} {2014})}\BibitemShut
  {NoStop}%
\bibitem [{xu2()}]{xu2014}%
  \BibitemOpen
  \href@noop {} {}\bibinfo {note} {X. Xu, private communication.}\BibitemShut
  {Stop}%
\bibitem [{Note3()}]{Note3}%
  \BibitemOpen
  \bibinfo {note} {The in-pane field could also induce a valley current
  proportional to the anomalous velocity.~\cite {xiao2007,xiao2012,mak2014} To
  distinguish it from the valley photocurrent, one can use the property that
  the former (latter) is odd (even) under the reversal of the electric
  field.}\BibitemShut {Stop}%
\bibitem [{\citenamefont {Werake}\ and\ \citenamefont
  {Zhao}(2010)}]{werake2010}%
  \BibitemOpen
  \bibfield  {author} {\bibinfo {author} {\bibfnamefont {L.~K.}\ \bibnamefont
  {Werake}}\ and\ \bibinfo {author} {\bibfnamefont {H.}~\bibnamefont {Zhao}},\
  }\href@noop {} {\bibfield  {journal} {\bibinfo  {journal} {Nature Phys.}\
  }\textbf {\bibinfo {volume} {6}},\ \bibinfo {pages} {875} (\bibinfo {year}
  {2010})}\BibitemShut {NoStop}%
\bibitem [{\citenamefont {Wang}\ \emph {et~al.}(2010)\citenamefont {Wang},
  \citenamefont {Zhu},\ and\ \citenamefont {Liu}}]{wang2010}%
  \BibitemOpen
  \bibfield  {author} {\bibinfo {author} {\bibfnamefont {J.}~\bibnamefont
  {Wang}}, \bibinfo {author} {\bibfnamefont {B.-F.}\ \bibnamefont {Zhu}}, \
  and\ \bibinfo {author} {\bibfnamefont {R.-B.}\ \bibnamefont {Liu}},\ }\href
  {\doibase 10.1103/PhysRevLett.104.256601} {\bibfield  {journal} {\bibinfo
  {journal} {Phys. Rev. Lett.}\ }\textbf {\bibinfo {volume} {104}},\ \bibinfo
  {pages} {256601} (\bibinfo {year} {2010})}\BibitemShut {NoStop}%
\bibitem [{\citenamefont {Glazov}\ and\ \citenamefont
  {Ganichev}(2014)}]{glazov2014}%
  \BibitemOpen
  \bibfield  {author} {\bibinfo {author} {\bibfnamefont {M.~M.}\ \bibnamefont
  {Glazov}}\ and\ \bibinfo {author} {\bibfnamefont {S.~D.}\ \bibnamefont
  {Ganichev}},\ }\href@noop {} {\bibfield  {journal} {\bibinfo  {journal}
  {Phys. Rep.}\ }\textbf {\bibinfo {volume} {535}},\ \bibinfo {pages} {101}
  (\bibinfo {year} {2014})}\BibitemShut {NoStop}%
\bibitem [{\citenamefont {Yu}\ \emph {et~al.}(2014)\citenamefont {Yu},
  \citenamefont {Wu}, \citenamefont {Liu}, \citenamefont {Xu},\ and\
  \citenamefont {Yao}}]{yu2014}%
  \BibitemOpen
  \bibfield  {author} {\bibinfo {author} {\bibfnamefont {H.}~\bibnamefont
  {Yu}}, \bibinfo {author} {\bibfnamefont {Y.}~\bibnamefont {Wu}}, \bibinfo
  {author} {\bibfnamefont {G.-B.}\ \bibnamefont {Liu}}, \bibinfo {author}
  {\bibfnamefont {X.}~\bibnamefont {Xu}}, \ and\ \bibinfo {author}
  {\bibfnamefont {W.}~\bibnamefont {Yao}},\ }\href {\doibase
  10.1103/PhysRevLett.113.156603} {\bibfield  {journal} {\bibinfo  {journal}
  {Phys. Rev. Lett.}\ }\textbf {\bibinfo {volume} {113}},\ \bibinfo {pages}
  {156603} (\bibinfo {year} {2014})}\BibitemShut {NoStop}%
\bibitem [{\citenamefont {{Muniz}}\ and\ \citenamefont
  {{Sipe}}(2014)}]{sipe2014}%
  \BibitemOpen
  \bibfield  {author} {\bibinfo {author} {\bibfnamefont {R.~A.}\ \bibnamefont
  {{Muniz}}}\ and\ \bibinfo {author} {\bibfnamefont {J.~E.}\ \bibnamefont
  {{Sipe}}},\ }\href@noop {} {\bibfield  {journal} {\bibinfo  {journal}
  {arXiv:1409.2555}\ } (\bibinfo {year} {2014})}\BibitemShut {NoStop}%
\bibitem [{\citenamefont {Mak}\ \emph {et~al.}(2013)\citenamefont {Mak},
  \citenamefont {He}, \citenamefont {Lee}, \citenamefont {Lee}, \citenamefont
  {Hone}, \citenamefont {Heinz},\ and\ \citenamefont {Shan}}]{mak2013}%
  \BibitemOpen
  \bibfield  {author} {\bibinfo {author} {\bibfnamefont {K.~F.}\ \bibnamefont
  {Mak}}, \bibinfo {author} {\bibfnamefont {K.}~\bibnamefont {He}}, \bibinfo
  {author} {\bibfnamefont {C.}~\bibnamefont {Lee}}, \bibinfo {author}
  {\bibfnamefont {G.~H.}\ \bibnamefont {Lee}}, \bibinfo {author} {\bibfnamefont
  {J.}~\bibnamefont {Hone}}, \bibinfo {author} {\bibfnamefont {T.~F.}\
  \bibnamefont {Heinz}}, \ and\ \bibinfo {author} {\bibfnamefont
  {J.}~\bibnamefont {Shan}},\ }\href@noop {} {\bibfield  {journal} {\bibinfo
  {journal} {Nature Mater.}\ }\textbf {\bibinfo {volume} {12}},\ \bibinfo
  {pages} {207} (\bibinfo {year} {2013})}\BibitemShut {NoStop}%
\bibitem [{\citenamefont {Liu}\ \emph {et~al.}(2014)\citenamefont {Liu},
  \citenamefont {Ansah~Antwi}, \citenamefont {Chua},\ and\ \citenamefont
  {Chi}}]{liu2014}%
  \BibitemOpen
  \bibfield  {author} {\bibinfo {author} {\bibfnamefont {H.}~\bibnamefont
  {Liu}}, \bibinfo {author} {\bibfnamefont {K.~K.}\ \bibnamefont
  {Ansah~Antwi}}, \bibinfo {author} {\bibfnamefont {S.}~\bibnamefont {Chua}}, \
  and\ \bibinfo {author} {\bibfnamefont {D.}~\bibnamefont {Chi}},\ }\href@noop
  {} {\bibfield  {journal} {\bibinfo  {journal} {Nanoscale}\ }\textbf {\bibinfo
  {volume} {6}},\ \bibinfo {pages} {624} (\bibinfo {year} {2014})}\BibitemShut
  {NoStop}%
\bibitem [{\citenamefont {Cheiwchanchamnangij}\ and\ \citenamefont
  {Lambrecht}(2012)}]{cheiwchanchamnangij2012}%
  \BibitemOpen
  \bibfield  {author} {\bibinfo {author} {\bibfnamefont {T.}~\bibnamefont
  {Cheiwchanchamnangij}}\ and\ \bibinfo {author} {\bibfnamefont {W.~R.~L.}\
  \bibnamefont {Lambrecht}},\ }\href {\doibase 10.1103/PhysRevB.85.205302}
  {\bibfield  {journal} {\bibinfo  {journal} {Phys. Rev. B}\ }\textbf {\bibinfo
  {volume} {85}},\ \bibinfo {pages} {205302} (\bibinfo {year}
  {2012})}\BibitemShut {NoStop}%
\bibitem [{\citenamefont {Radisavljevic}\ \emph {et~al.}(2011)\citenamefont
  {Radisavljevic}, \citenamefont {Radenovic}, \citenamefont {Brivio},
  \citenamefont {Giacometti},\ and\ \citenamefont {Kis}}]{radisavljevic2011}%
  \BibitemOpen
  \bibfield  {author} {\bibinfo {author} {\bibfnamefont {B.}~\bibnamefont
  {Radisavljevic}}, \bibinfo {author} {\bibfnamefont {A.}~\bibnamefont
  {Radenovic}}, \bibinfo {author} {\bibfnamefont {J.}~\bibnamefont {Brivio}},
  \bibinfo {author} {\bibfnamefont {V.}~\bibnamefont {Giacometti}}, \ and\
  \bibinfo {author} {\bibfnamefont {A.}~\bibnamefont {Kis}},\ }\href@noop {}
  {\bibfield  {journal} {\bibinfo  {journal} {Nature Nanotech.}\ }\textbf
  {\bibinfo {volume} {6}},\ \bibinfo {pages} {147} (\bibinfo {year}
  {2011})}\BibitemShut {NoStop}%
\end{thebibliography}
\end{document}